\documentclass{kluwer}    
\usepackage[dvips]{graphicx}
\newdisplay{guess}{Conjecture}

\begin{document}
\begin{article}
\begin{opening}
\title{Status of the connection between unidentified EGRET sources and
supernova remnants: The case of CTA~1}
\author{Diego F. \surname{Torres}$^1$, Thomas M.  \surname{Dame}$^2$
\& Gustavo E. \surname{Romero}$^3$ } \runningauthor{Torres, Dame,
Romero} \runningtitle{SNRs and Unidentified EGRET sources}
\institute{$^1$Lawrence Livermore
Laboratory, 7000 East Ave. L-413, Livermore, CA 94550, USA\\
$^2$ Harvard-Smithsonian Center for Astrophysics, 60 Garden
Street, Cambridge, MA 02138, USA\\
$^3$Instituto Argentino de Radioastronom\'{\i}a, C.C.5, 1894 Villa
Elisa, Buenos Aires, Argentina}
\date{}

\begin{abstract} In this paper we briefly comment on the observational
status of the possible physical association between unidentified
EGRET sources and supernova remnants (SNRs) in our Galaxy. We draw
upon recent results presented in the review by Torres et al.
(Physics Reports, 2003), concerning molecular gas in the vicinity of
all 19 SNRs found to be positionally coincident with EGRET sources
at low Galactic latitudes. In addition, we present new results
regarding the supernova remnant CTA~1. Our findings disfavor the
possibility of a physical connection with the nearby (in projection)
EGRET source. There remains possible, however, that the compact
object produced in the supernova explosion be related with the
observed $\gamma$-ray flux.

\end{abstract}


\end{opening}

\section{A brief overview of the EGRET--SNR connection}

We have recently presented a case-by-case analysis of the positional
correlation between unidentified EGRET sources and supernova
remnants (SNRs, Torres et al. 2003). There are 19 EGRET sources at
low Galactic latitudes ($|b|<10^o$) in spatial correlation with,
mostly shell-type, SNRs. The Poisson probability for the 19
coincidences to be a chance effect is $1.05 \times 10^{-5}$, i.e.,
there is an a priori 0.99998 probability that at least one of the
positional associations is physical. This expected chance
association was computed using a numerical code described by Romero
et al. (1999), and Sigl et al. (2001). With the aim to identify the
most plausible candidates for an EGRET--SNR connection, we have
analyzed, among other things, the variability status of the EGRET
sources and the potential contribution from known radio pulsars to
the high energy flux (Torres et al. 2003).

In order to understand the origin of all the unidentified
detections, their variability status is of paramount importance
(Nolan 2003, Torres et~al. 2001). Classic known models for
$\gamma$-ray sources in our Galaxy, like a SNR--molecular cloud
interaction (e.g. Aharonian \& Atoyan 1996), would produce
non-variable detections during the timescale of EGRET observations.
To classify the variability status we take into account both indices
quoted in the literature, Torres et~al.'s (2001) $I$ and Tompkins'
$\tau$. They are different in nature, although results have proven
to be statistically correlated (Torres, Pessah \& Romero 2001). The
index $I$ is a relative classification of variability with respect
to the pulsar population. Sources with $I>2.5$ are more than
3$\sigma$ away from the statistical variability of pulsars.
Similarly, sources for which $\tau$ is at least 0.6, are also more
than $3\sigma$ away from the mean value of the $\tau$ upper limit
for pulsars, and are thus classified as variable. For comparison,
note that the mean value of $\tau$ ($I$) for known AGNs is 0.9
(3.3). We find that most EGRET sources in positional coincidence
with SNRs, and certainly those which represent the best cases for a
physical association, are non-variable.

In order to unambiguously identify a pulsar as the origin of the
$\gamma$-rays from an EGRET source, $\gamma$-ray pulsations must be
detected at the pulsar period.  However the $\gamma$-fluxes are
generally very low, and nearly contemporaneous radio/X-ray and
$\gamma$-ray observations are required to fold the few available
photons with the correct ephemeris.  Because this is no longer
possible owing to the demise of the Compton Gamma-Ray Observatory,
new candidate associations must mostly be judged by comparison with
the properties of the six known EGRET pulsars and those of the known
best candidates. We have done so for every known radio pulsar found
to be coincident with these 19 EGRET sources, and have isolated
several interesting candidates for an EGRET source--pulsar
connection. These are mostly found with young pulsars pertaining to
the recently released Parkes Multibeam Survey (see, e.g. Torres,
Butt, \& Camilo 2001, Torres \& Nuza 2003). This scenario, then, is
a competing one to the EGRET--SNR connection and has to be taken
into account in judging the plausibility of the latter.

The need of a case-by-case analysis is then clearly established.
Several cases for a physical association, based on the analysis of
$\gamma$-ray data, the molecular environment of the SNR, and the
analysis of any coincident pulsar (if any), have been isolated and
reviewed in Torres et al. (2003). Among the most notable cases for
such a physical association are G347.3-0.5, W66, W28 and IC443. The
first of these, providing probably the best evidence to date for a
hadronic origin of the $\gamma$-ray source detected by EGRET, has
received much attention recently (Butt et al. 2001, 2002, Enomoto
2002, Uchiyama et al. 2002). In addition, the case-by-case analysis
shows that it is at least plausible, contrary to expectations, that
EGRET has detected distant (more than 6 kpc) SNRs. There are 5
coinciding pairs of 3EG sources and SNRs for which the latter
apparently lie at such high values of distance (disregarding those
related with SNRs spatially close to the galactic center). For all
these cases, we have uncovered the existence of nearby, large, in
some cases giant, molecular clouds that could enhance the GeV signal
through pion decay. It is possible that the physical relationship
between the 3EG source and the coincident SNR could provide for
these pairs a substantial part of the GeV emission observed. This
does not preclude, however, composite origins for the total amount
of the radiation detected, since some of these cases present other
plausible scenarios (e.g. energetic pulsars in the EGRET field).
AGILE observations, in advance of GLAST, would greatly elucidate the
origin for these 3EG sources, since even a factor of 2 of
improvement in spatial resolution would be enough to reject the
EGRET--SNR connection.

In what follows we attempt to briefly analyze, similarly to what was
done in the case-by-case analysis of Torres et al. (2003), the pair
CTA~1/3EG J0010+7309, a supernova remnant lying at relatively high
Galactic latitudes, and for which there has been little analysis on
their CO environment.

\section{CTA~1: General Features}

CTA 1 is a shell-type SNR, but its shell is incomplete and
broken-out towards the NW. This breakout phenomenon may be caused by
more rapid expansion of the blast wave shock into a lower density
region toward the NW. HI observations supported this interpretation
(Pineault et al. 1993, 1997). The latter papers place the SNR at
1.4$\pm$0.3 kpc, the kinematic distance corresponding to the
systemic velocity of a partial HI shell ($-16$ km/s) that the
authors find to the NW. However, it is to be noticed that the
systemic (unperturbed) velocity of a partially expanding shell might
be subject to a great uncertainty.

Since CTA 1 is located at a relatively high latitude ($b=$10$^o$)
and it is nearby, it has a large angular size (90 arcmin), little
foreground or background confusion, and it can be observed at
exceptionally high linear resolution. The age of the SNR was
estimated to be 10$^4$ yr by Pineault et al. (1993), but it could be
younger by a factor of 2 (Slane et al. 1997). CTA 1 was subject of
intense observational campaigns in the past years. There have been
both ROSAT and ASCA X-ray observations (Seward et al. 1995, Slane et
al. 1997), as well as optical, infrared and radio studies (see
Pineault et al. 1997 and Brazier et al. 1998 for a review).

CTA 1 belongs to the class of composite SNRs (SNRs with central
pulsar wind nebulae), which show a shell-type morphology in the
radio band and are center-filled in X-rays. Five point sources were
detected with ROSAT, one of which was found to coincide with the
EGRET source (at the time of the analysis, 2EG J0008+7307), see
Figure 3 of Brazier et al. (1998). ASCA data suggested that this
source, named RX J0007.0+7302, has a non-thermal spectrum, and gave
support to the idea that it is the pulsar left from the supernova
explosion (Slane et al. 1997). Recent observations by Slane et al.
(2004), using XMM, confirm that the X-ray spectrum of the source is
consistent with that of a neutron star and is well described by a
power law with the addition of a soft thermal component that may
correspond to emission from hot polar cap regions or to cooling
emission from a light-element atmosphere over the entire star. An
extrapolation of the non-thermal spectrum of RX J0007.0+7302 to
$\gamma$-ray energies would yield a flux consistent with that of
EGRET source 3EG J0010+7309, supporting the proposition that these
sources could be related. A Chandra image of the central X-ray
source RX J0007.0+7303 (Halpern et al. 2004) reveals a point source,
a compact nebula, and a bent jet, all of which are characteristic of
energetic, rotation-powered pulsars.

Optical observations were carried out by Brazier et al. (1998), with
a 2.12-m telescope, but no object was found within the positional
error box of the X-ray source. This allowed an upper limit to be set
on the optical magnitude of any counterpart to the putative pulsar.
Halpern et al. (2004) also obtained upper limits in the optical at
the position of the point source, corresponding to an
X-ray-to-optical flux ratio larger than 100. Neither a VLA image at
1425 MHz nor a deep pulsar search at 820 MHz using the NRAO Green
Bank Telescope, revealed a radio pulsar counterpart

The $\gamma$-ray source 3EG J0010+7309 is non-variable under the $I$
and $\tau$ schemes, and has a hard spectral index of 1.85$\pm$0.10,
compatible with those of the Vela pulsar. This source was also
detected in the Second EGRET Catalog, but with a shifted position
which made it coincide with a nearby AGN -- which was at the time
proposed as a possible counterpart (Nolan et al. 1996). This AGN is
not coincident with the 3EG source and can not be considered a
plausible counterpart any longer. Based on positional coincidence,
on the hard spectral index, and on physical similarities between the
Vela pulsar and RX J0007.0+7302, Brazier et al. (1998) proposed that
the 3EG source and this X-ray source were related. For an assumed 1
sr beaming, the observed 100-2000 MeV flux corresponds to a
luminosity of $4\times 10^{33}$ erg s$^{-1}$, compatible with other
$\gamma$-ray pulsar detections. A competing explanation to the
$\gamma$-ray pulsar would be given if a massive and dense molecular
cloud is found to be interacting with, or being overtaken by, the
supernova remnant shock. The molecular environment of CTA~1 is
analyzed below to further assess this possibility.

As can be noted from the figure, the EGRET source seems too far from
the SNR nearest rim.  However, if some SNRs interact, as expected,
with nearby massive clouds producing enhanced $\gamma$-ray emission
through hadronic/Bremsstrahlung interactions, cases in which there
is just a marginal coincidence between the SNRs and the centers of
the EGRET sources should be also considered, since the peak
$\gamma$-ray emissivity will likely be biased towards the adjacent
cloud.

\section{CTA~1: Molecular environment}

As discussed in Dame et al. (1987), much of the local (closer than
Perseus Arm) CO emission in the second quadrant of the Galaxy
bifurcates into 2 velocity components, one centered near $-12$ km/s
and the other close to 0 km/s. Dame et al. (1987) argue that the
``$-12$ km/s clouds" are probably associated with a string of OB
associations at 800 to 1000 pc from Earth. The clouds with
velocities near  0 km/s were associated  with the so-called
``Lindblad Ring" at ~300 pc. The long thin cloud partially
overlapping CTA 1 (see Figure \ref{co}) has a mean velocity of
$\sim-3$~km/s, suggesting that it is part of the Lindblad Ring
population. In fact, these clouds appear to be part of a very large
loop of local clouds (all in the range $-5$ to 0 km/s). Grenier et
al. (1989) proposed that this loop surrounds a $4 \times 10^4$ yr
old Type I SNR 300 pc away. Thus, the velocity of the cloud near to
CTA 1 suggests that it is part of this larger system, lying at only
$\sim $300 pc.

\begin{figure}[t]
 \caption{Molecular gas in the direction of the SNR
  CTA 1.  (a) The
  false-color map is CO integrated over the velocity range
  $-15$ to $+5$
  km/s.  The black contours are 4.85 GHz intensity from
  the survey of
  Condon et al. (1994), which show CTA 1 as a
  semi-circular feature
  near the center of the map.  The red (lighter) contours are the
  50\%, 68\%, 95\%,
  and 99\% confidence contours for the unidentified gamma
  ray source 3EG
  J0010+7309.
  (b) A CO latitude-velocity map integrated from 116$^o$ to
  120$^o$  in
  Galactic longitude.
[[The figure is available on-line as a jpeg file]]}\label{co}
\end{figure}

The total $H_2$ mass (calculated from CO data) of the long thin
cloud near CTA1 (in the region $l=116^o$ to 120$^o$, $b=8^o$ to
13$^o$, $v=-15$ to 5 km/s) is 4900 $M_\odot$, assuming a distance of
300 pc. However, there is no evidence of interaction between the
remnant and the cloud (what is consistent with the different most
plausible distances to each of these objects). Figure \ref{co}b
attached, for example, shows a latitude--velocity map integrated
over the longitude extent of the cloud. There is little indication
of accelerated gas at the latitude/velocity of the remnant, $b =9^o$
to 10$^o$ and $\sim -15$ km/s, respectively. The cloud may well be,
then, in the foreground, and unrelated to CTA 1. In addition to
this, there is no CO emission detected within the contours of the
3EG source, which also argues against a hadronic origin of the
$\gamma$-ray source.

Pineault et al.'s (1993) systemic velocity of $-16$ km/s for the HI
shell is reasonable close to that of the ``$-12$ km/s clouds", which
in turn are probably associated with the OB associations Cep OB2,
Cep OB3, Cep OB4, and Per OB3 at $\sim 800$ pc. This might be
suggesting that the latter is the most likely distance for CTA 1,
although it remains uncertain.
In any case, it seems that CTA 1 and 3EG J0010+7309 might be
physically related only through the compact object left by the
supernova explosion, RX J0007.0+7302. A better localization of the
$\gamma$-ray source with AGILE and GLAST will certainly test this
suggestion. For the moment, our CO analysis shows that a molecular
cloud origin of the $\gamma$-ray source  3EG J0010+7309 (for
instance, through hadronic processes similar to the case of
G347.3-0.5 proposed by Butt et al. 2001) is untenable.

\section*{Acknowledgments}
The work of DFT was performed under the auspices of the U.S.
Department of Energy (NNSA) by University of California Lawrence
Livermore National Laboratory under contract No. W-7405-Eng-48.
G.E.R. was supported by CONICET (under grant PIP N$^o$ 0430/98),
ANPCT (PICT 03-04881), as well as by Fundaci\'on Antorchas. We thank
P. Slane for comments.

\end{article}

\begin{thebibliography}{999}


\bibitem{} Brazier K.T.S., Reimer O.,  Kanbach G., \&
  Carrami\~nana A. 1998, MNRAS 295, 819

\bibitem{Butt:2001ff}
Butt Y., et al. 2001, ApJ 562, L167

\bibitem{Butt:2002im}
Butt Y., et al. 2002, Nature 418, 499

\bibitem{} Condon J.J., et al. 1994, AJ 107, 1829



\bibitem{} Dame T.M., et al. 1987, ApJ 322, 706

\bibitem{} Enomoto R., et al. 2002, Nature 416, 823
\bibitem{} Grenier I. et al. 1989, ApJ 347, 231

\bibitem{}  Halpern J. P., Gotthelf E. V., Camilo F.,  Helfand D. J. \&
Ransom S. M. 2004, astro-ph/0404312

\bibitem{} Nolan P.L., et al. 1996, ApJ 459, 100
\bibitem{} Nolan P.L., Tompkins W.F., Grenier I.A., \& Michelson
P.F. 2003, ApJ 597, 615

\bibitem{} Pineault S., et al. 1993, AJ 105, 1060
\bibitem{} Pineault S., et al. 1997, A\&A 324, 1152


\bibitem{} Slane P., Zimmerman E. R., Hughes J. P., Seward
F. D., Gaensler B. M. \& Clarke M. J. 2004, ApJ 601, 1045

\bibitem{Romero:1999tk}
Romero G.E., Benaglia P., \& Torres D.F. 1999, A\&A 348, 868



\bibitem{} Seward F.D., Schmidt B., \& Slane P. 1995, ApJ 453, 284

\bibitem{Sigl:2000sn}
Sigl G., et al. 2001, Phys. Rev. D63, 081302

\bibitem{} Slane P., et al. 1997, ApJ 485, 221

\bibitem{Torres:xa}
Torres D.F., et al.  2001, A\&A 370, 468

\bibitem{Torres:zu}
Torres D.F., Pessah M.E., \& Romero G.E. 2001c, Astronomische Nach.
322, 223

\bibitem{Torres:2001zb}
 Torres D.F., Butt Y.M. \& Camilo F. 2001, ApJ 560, L155


\bibitem{Torres:2002rs}
Torres D.F., \& Nuza S. E. 2003, ApJ 583, L25

\bibitem{Torres:2002af}
Torres D.F., et al.  2003, Physics Reports 382, 303

\bibitem{} Uchiyama Y., Takahashi T., \& Aharonian F.A. 2002a,
PASJ54, L73
\end{thebibliography}
\end{document}